\documentclass[letterpaper,preprintnumbers,english,floats,floatfix,amssymb,prd,superscriptaddress,onecolumn,nofootinbib]{revtex4}

\usepackage{amssymb,amsmath}
\usepackage{epsfig,hyperref}
\usepackage{epstopdf}
\usepackage[british]{babel}
\usepackage{enumerate}
\usepackage{enumitem}
\usepackage{color}

\def\be{\begin{equation}}
\def\ee{\end{equation}}
\def\beq{\begin{eqnarray}}
\def\eeq{\end{eqnarray}}

\makeatletter

\newcommand{\arXiv}[2][]{\href{http://arxiv.org/abs/#2}{\texttt{arXiv:#2\@ifempty{#1}{}{ [#1]}}}}
\makeatother

\begin{document}
\title{New results on the dynamics of critical collapse}

\author{Jun-Qi Guo}%
\email{sps{\_}guojq@ujn.edu.cn}
\affiliation{School of Physics and Technology, University of Jinan, Jinan 250022, Shandong, China}

\author{Yu Hu}%
\email{yuhu@hust.edu.cn}
\affiliation{MOE Key Laboratory of Fundamental Physical Quantities Measurement, Hubei Key Laboratory of Gravitation and Quantum Physics, PGMF, and School of Physics, Huazhong University of Science and Technology, Wuhan 430074, Hubei, China}

\author{Pan-Pan Wang\textsuperscript{2}}%
\email{ppwang@hust.edu.cn}

\author{Cheng-Gang Shao\textsuperscript{2}}%
\email{cgshao@hust.edu.cn}

\date{\today}

\begin{abstract}
We study the dynamics of the critical collapse of a spherically symmetric scalar field. Approximate analytic expressions for the metric functions and matter field in the large-radius region are obtained. In the central region, owing to the boundary conditions, the equation of motion for the scalar field is reduced to the flat-spacetime form.
\end{abstract}
\maketitle

\section{Introduction}
The critical phenomena in gravitational collapse discovered by Choptuik demonstrate the rich dynamics of the Einstein equations~\cite{Choptuik:1992jv}. Consider the gravitational collapse of generic families of a massless scalar field, whose initial data are parameterized by $p$. The parameter $p$ measures the strength of the gravitational interaction. Strong interactions (high $p$) result in black hole formation, whereas weak interactions (low $p$) disperse the matter field to infinity and flat spacetime remains. By fine-tuning $p$ to the threshold of black hole formation, $p=p_*$, critical collapse occurs.

In supercritical collapse, a tiny black hole forms, the mass of which has the scaling relation, $m_{BH}\propto|p-p_*|^{\gamma}$, where $\gamma\simeq 0.37$. The critical collapse solution exhibits universality feature, namely, the spacetime produced by different families of critical initial data approach to the same solution after a finite time. The solution also displays discrete self-similarity: it is invariant after rescaling the spacetime by a certain factor.
Since this discovery, similar results have been obtained in many other models (see Ref.~\cite{Gundlach:2007gc} for a review). The critical behavior in the 3D scalar collapse was studied in Ref.~\cite{Deppe:2018uye}. The critical phenomena in the collapse of electromagnetic waves were simulated in Refs.~\cite{Baumgarte:2019fai,Mendoza:2021nwq}. The interplay between multiple near-critical fields in spherical scalar collapse was investigated in Ref.~\cite{Kelson-Packer:2020hbb}. The critical phenomena in the bald/scalarized black hole phase transition occurring in Einstein-Maxwell-scalar theory were reported in Ref.~\cite{Zhang:2021nnn}. In simulations, near the threshold, all the intermediate solutions are attracted to the critical solution and remain in this state for a long time. The system resembles the behavior of quasinormal modes. In the late stage, the intermediate solutions decay into Reissner-Nordstr\"{o}m black holes for subcritical cases or scalarized charged black holes for supercritical cases, resembling the quasinormal modes.

Analytic interpretations are important for understanding the dynamics of gravitational collapse. In Refs.~\cite{Gundlach:1995kd,Gundlach:1996eg,Martin-Garcia:2003xgm}, critical collapse was treated as an eigenvalue problem. By imposing discrete self-similarity, the global structure of the critical collapse spacetime was constructed with the  pseudo-Fourier method. The rescaling factor $\Delta$ became an eigenvalue and was solved with high precision. The scaling law of the black hole mass in supercritical collapse was recovered analytically via the perturbation approach in Ref.~\cite{Gundlach:1996eg}. Critical collapse was analyzed using a renormalization group method in Refs.~\cite{Koike:1995jm,Hara:1996mc}. In Ref.~\cite{Reiterer:2012hnr}, using an explicit approximate solution, a true solution was shown to exist. In Ref.~\cite{Guo:2018yyt}, using a typical log-periodic formula in discrete scale invariance systems, the authors obtained an approximate analytic solution for the spacetime near the center. Approximate analytic expressions for the metric functions and matter field near the central singularity in black hole formation were obtained in Refs.~\cite{Guo:2013dha,Guo:2020jfa}. In Ref.~\cite{Price:1996sk}, the equations for the matter field in critical collapse were analyzed with certain terms in the equations dropped. Furthermore, approximate expressions for certain combinations of the metric functions and derivatives of the scalar field were obtained.

In this study, considering the significance of analytic results, we obtain approximate analytic expressions for the metric functions and matter field in the large-radius region using numerical data. We also investigate the dynamics in the central region. We find that owing to the boundary conditions at the center, the equation of motion for the scalar field in the central region is reduced to the flat-spacetime form.

This paper is organized as follows. In Sec.~\ref{sec:methodology}, we describe the methodology for simulating critical collapse. In Secs.~\ref{sec:large_radius} and \ref{sec:central}, we study the dynamics in the large-radius and central regions, respectively. Finally, the results are summarized in Sec.~\ref{sec:summary}.

\section{Methodology\label{sec:methodology}}
We simulate the critical collapse of a spherically symmetric massless scalar field $\phi$ in polar coordinates,
\be ds^{2}=-A(r,t)\mbox{e}^{-2\delta(r,t)}\mbox{d}t^{2}+\frac{1}{A(r,t)}\mbox{d}r^{2}+r^{2}\mbox{d}\Omega^2.\label{metric1}\ee
The energy-momentum tensor for the scalar field is
$T_{\mu\nu}=\phi_{,\mu}\phi_{,\nu}-(1/2)g_{\mu\nu}g^{\alpha\beta}\phi_{,\alpha}\phi_{,\beta}$.
Some components of the Einstein and energy-momentum tensors for $\phi$ are shown below.
\be G^{t}_{t}=\frac{1}{r^2}(rA_{,r}-1+A),\hphantom{dddd} G^{r}_{r}=-\frac{1}{r^2}(-rA_{,r}+2rA\delta_{,r}+1-A),\hphantom{dddd} G^{r}_{t}=-\frac{1}{r}A_{,t},
\nonumber\ee
\be T^{t}_{t}=-T^{r}_{r}=-\frac{1}{2}A(P^2+Q^2),\hphantom{dddd} T^{r}_{t}=A\phi_{,r}\phi_{,t}.\nonumber\ee
Here, $(_{,r})$ and $(_{,t})$ denote partial derivatives with respect to the coordinates $r$ and $t$, respectively.

We define
\be Q(r,t)\equiv\phi_{,r}, \hphantom{ddd} P(r,t)\equiv A^{-1}\mbox{e}^{\delta}\phi_{,t},\label{define_Q_P}\ee
and set $G=1$. Then, the equations $G^{t}_{t}=8\pi T^{t}_{t}$ and $G^{r}_{t}=8\pi T^{r}_{t}$ respectively generate
\be A_{,r}=\frac{1-A}{r}-4\pi rA(P^{2}+Q^{2}),\label{eomA}\ee
\be A_{,t}=-8 \pi rA^{2}\mbox{e}^{-\delta}PQ.\label{eqAt}\ee
The combination of $G^{r}_{r}=8\pi T^{r}_{r}$ and Eq.~(\ref{eomA}) yields
\be \delta_{,r}=-4\pi r(P^{2}+Q^{2}).\label{eomd}\ee
With Eq.~(\ref{define_Q_P}), we obtain
\be Q_{,t}=(A \mbox{e}^{-\delta} P)_{,r}.\label{eomQ}\ee
From the conservation of the energy-momentum tensor, $T^{\mu\nu}_{;\mu}=0$, we have
\be P_{,t}=\frac{1}{r^{2}}(r^{2} A \mbox{e}^{-\delta} Q)_{,r}.\label{eomP}\ee
The Misner-Sharp mass is defined as~\cite{Misner_1964}
\be m\equiv\frac{r}{2}(1-g^{\mu\nu}r_{,\mu}r_{,\nu})=\frac{r}{2}(1-A).\label{MS_mass}\ee

The initial conditions for $\phi$ are set as $\phi|_{t_i}=a\exp[-(r/\sigma)^{2}]$ and $\phi_{,t}|_{t_i}=0$. The regularity of Eq.~\eqref{eomA} at the center requires that $A(r=0,t)=1$. We choose $\delta(r=0,t)=0$, which implies that the coordinate time is equal to its proper time at the center. In the simulation, we integrate Eqs.~(\ref{eomA}) and (\ref{eomd})-(\ref{eomP}) using the fourth-order Runge-Kutta method. A mesh refinement algorithm is implemented. We approach to the critical solution by fine-tuning the initial profile of the scalar field via the bisection method. For details on the numerics, see Ref.~\cite{Zhang:2016kzg}.

\section{Result I: dynamics in the large-radius region\label{sec:large_radius}}
We rewrite the metric~(\ref{metric1}) as
\be ds^{2}=-\alpha^{2}(r,t)dt^{2}+\beta^{2}(r,t)dr^{2}+r^{2}d\Omega^2.\ee
For convenience, we adjust the time coordinate such that $t=0$ when the naked singularity forms, and
define $X(r,t)\equiv\sqrt{2\pi}(r/\beta)\phi_{,r}$, $Y(r,t)\equiv\sqrt{2\pi}(r/\alpha)\phi_{,t}$,
$\rho\equiv \ln r$, $T\equiv\ln(-t)$, and $u\equiv t/r$. Then, the equations for $\phi$~(\ref{eomQ}) and (\ref{eomP}) can be respectively rewritten as
\be (\beta X)_{,u}=-\alpha Y + (\alpha Y)_{,\rho} -u(\alpha Y)_{,u},\label{aX_u}\ee
\be (\beta Y)_{,u}=\alpha X + (\alpha X)_{,\rho}-u(\alpha X)_{,u}.\label{aY_u}\ee

In critical collapse, the period in terms of the coordinate time $t$ exponentially decreases. Consequently, the metric functions and matter field in the late stage of collapse and large-radius region for which $|t/r|\ll 1$ appear to be \lq\lq frozen\rq\rq rather than propagating~\cite{Choptuik_workshop_1993,Choptuik:1997mq}. In Ref.~\cite{Price:1996sk}, the authors offered one ansatz: in this region, the last terms in Eqs.~(\ref{aX_u}) and (\ref{aY_u}) are negligible in comparison with the first terms. Moreover, treating $\alpha$ and $\beta$ as constant, the authors obtained the following solutions:
\be
X\approx B\sin[\omega(\rho-\alpha u)-\gamma], \hphantom{ddd} Y\approx B\sin[\omega(\rho-\alpha u)],
\label{solution_XY}
\ee
where
\be 1+\frac{1}{\omega^{2}}=\beta^2, \hphantom{ddd} \sin\gamma=\frac{1}{\omega\beta}, \hphantom{ddd} \cos\gamma=-\frac{1}{\beta}.\ee
The expressions~(\ref{solution_XY}) are consistent with the numerical results. However, some treatments in the above have not been fully justified. In addition, although approximate expressions for $X$ and $Y$ were obtained, the results for the metric functions and scalar field remain absent. We address such issues below.

In Ref.~\cite{Price:1996sk}, some terms in Eqs.~(\ref{aX_u}) and (\ref{aY_u}), that is, $-u(\alpha Y)_{,u}$, $-u(\alpha X)_{,u}$, $\beta_{,u}X$, $\alpha_{,\rho}Y$, $\beta_{,u}Y$, and $\alpha_{,\rho}X$, were dropped. In fact, as shown in Figs.~\ref{fig:eom_X} and~\ref{fig:eom_Y}, in the large-radius region $(r>10^{-3})$, the absolute values of the terms $-u\alpha_{,u}Y$, $-u\alpha_{,u}X$, $\alpha_{,\rho}Y$, and $\alpha_{,\rho}X$ can sometimes be greater than those of other terms. However, the terms dropped approximately cancel. Consequently, the equations constructed by the remaining terms roughly hold:
\be \beta X_{,u}\approx-\alpha Y + \alpha Y_{,\rho},\label{eq_X}\ee
\be \beta Y_{,u}\approx\alpha X + \alpha X_{,\rho}.\label{eq_Y}\ee
From this perspective, treating $\alpha$ and $\beta$ as constant in Ref.~\cite{Price:1996sk} is effectively valid.

In the analytic results~(\ref{solution_XY}) obtained in Ref.~\cite{Price:1996sk}, only the phases in the sine functions are functions of $r$ and $t$, and the amplitudes are constant. We examine the numerical results for $\phi$ and find that besides the phase, the amplitude for $\phi$ is also a function of $r$ and $t$. We find that the field $\phi$ admits the following approximate expression:
\be \phi(r,t)\approx C_{1}(1+C_{2}[H(r,t)])\cos(\omega\ln r + C_{3}[H(r,t)] + \varphi_{0} ).\label{phi_sln}\ee
The quantity $[H(r,t)]$ has the following features:
\begin{enumerate}[fullwidth,itemindent=0em,label=(\roman*)]
  \item We first define $H(r,t)$ in the usual manner,
%  %
  \be H(r,t)\equiv\frac{\omega\alpha t}{r}=\omega A^{1/2}e^{-\delta}\frac{t}{r}.\ee
  Then, we have
  \be H_{,t}=\frac{\omega\alpha}{r}+\frac{\omega\alpha_{,t}t}{r}+\frac{\omega_{,t}\alpha t}{r}.\ee
  \item $[H(r,t)]$ is defined as
  \be [H(r,t)]\equiv\frac{{\bar{\omega}}{\bar{\alpha}} t}{r}.\ee
  Here, $\bar{\omega}$ and $\bar{\alpha}$ are constant values of $\omega(r,t)$ and $\alpha(r,t)$. This is related to the treatment of $\alpha$ and $\beta$ as constant in the discussions of Eqs.~(\ref{eq_X}) and (\ref{eq_Y}).
  \begin{figure*}[!htbp]
    \centering
    \includegraphics[width=\textwidth]{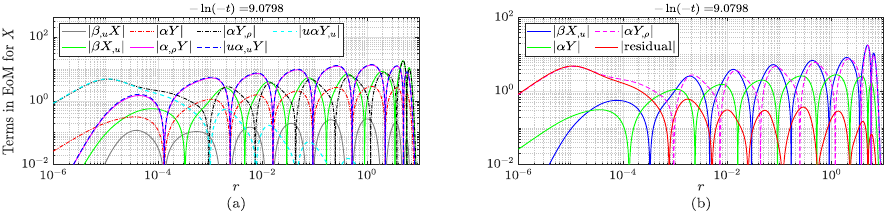}
    \caption{(color online) Numerical results for Eqs.~(\ref{aX_u}) and (\ref{eq_X}). (a) Results for Eq.~(\ref{aX_u}). (b) Results for Eq.~(\ref{eq_X}). In the large-radius region $(r>10^{-3})$, the absolute values of $\alpha_{,\rho}Y$ and $u\alpha_{,u}Y$ are sometimes greater than those of other terms in Eq.~(\ref{aX_u}). However, the three terms $\beta_{,u}X$, $-\alpha_{,\rho}Y$ and $u(\alpha Y)_{,u}$ roughly cancel. Then, Eq.~(\ref{eq_X}) approximately holds.}
    \label{fig:eom_X}
\end{figure*}

\begin{figure*}[!htbp]
    \centering
    \includegraphics[width=\textwidth]{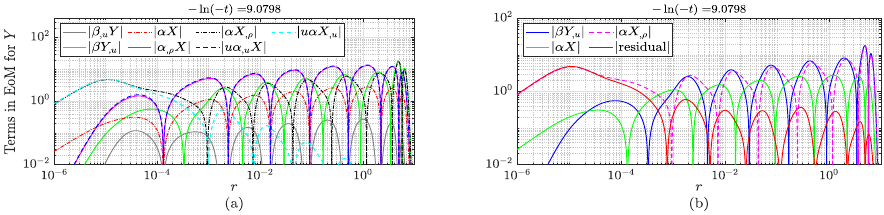}
    \caption{(color online) Numerical results for Eqs.~(\ref{aY_u}) and (\ref{eq_Y}). (a) Results for Eq.~(\ref{aY_u}). (b) Results for Eq.~(\ref{eq_Y}).}
    \label{fig:eom_Y}
\end{figure*}
  \item From Eq.~(\ref{phi_sln}), we obtain the expression for $\phi_{,t}$,
  \be \phi_{,t}\approx C_{1}\sqrt{C_{2}^2+C_{3}^2}[H]_{,t}\cos(\omega\ln r +C_{3}[H]+\varphi_{0}+\varphi_{1}),\label{phi_t}\ee
  where $\tan\varphi_{1}\equiv C_{3}/C_{2}$. Regarding the quantity $H_{,t}(=\omega\alpha/r+\omega\alpha_{,t}t/r+\omega_{,t}\alpha t/r)$, the numerical results show that $|{\omega\alpha_{,t}t}/{r}|$ is sometimes greater than ${\omega\alpha}/{r}$. However, comparing expression~(\ref{phi_t}) with the numerical results for $\phi_{,t}$, we always obtain
  \be [H]_{,t}\approx\frac{\omega\alpha}{r}=\omega A^{1/2}e^{-\delta}\frac{1}{r}.\label{H_t}\ee
  This implies that in $[H]_{,t}$, the contributions from $\omega\alpha_{,t}t/r$ and $\omega_{,t}\alpha t/r$ are negligible. This should be related to the fact that the respective reductions of Eqs.~(\ref{aX_u}) and (\ref{aY_u}) to Eqs.~(\ref{eq_X}) and (\ref{eq_Y}) are equivalent to treating $\alpha$ and $\beta(\equiv \sqrt{1+\omega^{-2}})$ as constant.
  \begin{figure*}[!htbp]
    \centering
    \includegraphics[width=\textwidth]{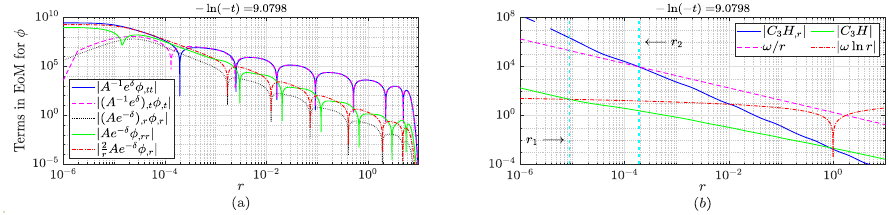}
    \caption{(color online) Numerical results for the equation of motion for $\phi$~(\ref{eomP}) and the transition region.
    (a) Results for Eq.~(\ref{eomP}). In the large-radius region $(r>10^{-3})$, $A^{-1}e^{\delta}\phi_{,tt}\approx-(A^{-1}e^{\delta})_{,t}\phi_{,t}$. This is very different from the flat-spacetime form $\phi_{,tt}=r^{-2}(r^{2}\phi_{,r})_{,r}$. In the central region $(r<10^{-5})$, $A^{-1}e^{\delta}\phi_{,tt}\approx Ae^{-\delta}r^{-2}(r^{2}\phi_{,r})_{,r}$. Considering that in this region $A\approx1$ and $\delta\approx 0$, $\phi_{,tt}\approx r^{-2}(r^{2}\phi_{,r})_{,r}$.
    (b) Transition region between the central and large-radius regions: $[r_1,~r_2]$. At $r=r_{1}$, $|C_{3}H|\sim|\omega\ln r|$. At $r=r_2$, $|C_{3}H_{,r}|\sim\omega/r$.}
    \label{fig:eom_phi}
\end{figure*}

\begin{figure*}[!htbp]
    \centering
    \includegraphics[width=0.85\textwidth]{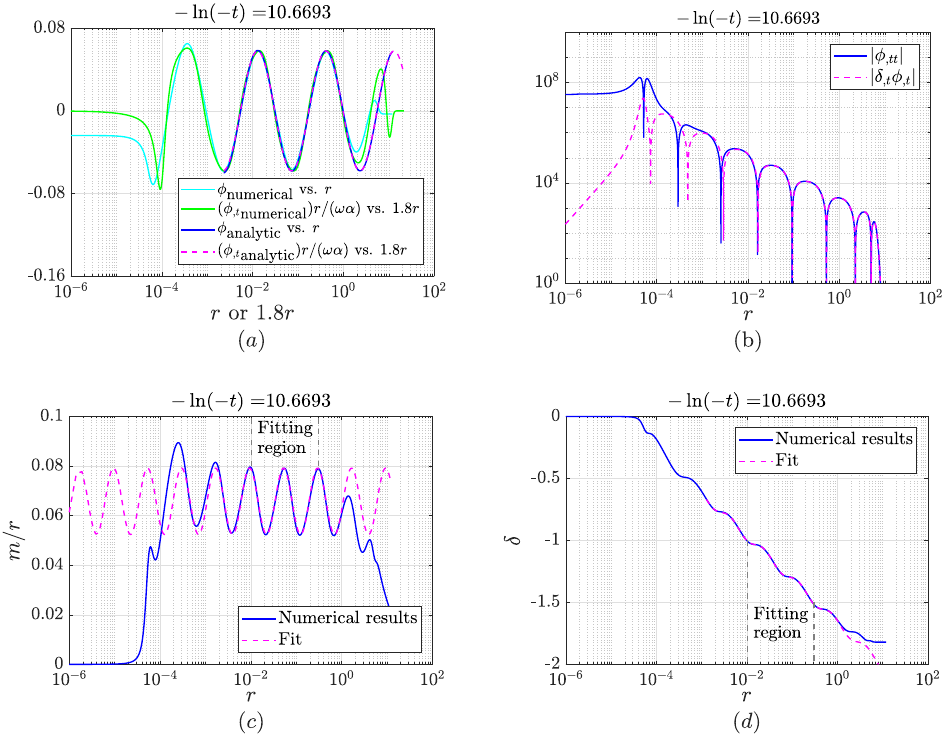}
    \caption{(color online) Comparison of the analytic and numerical results for the scalar field and metric functions in the large-radius region. (a) Results for $\phi$ (\ref{phi_sln}) and $\phi_{,t}$ (\ref{phi_t}). (b) $\phi_{,tt}\approx-\delta_{,t}\phi_{,t}$. (c) Results for $m/r$ (\ref{m_r_sln}). (d) Results for $\delta$ (\ref{delta_sln}).}
    \label{fig:phi_fit}
\end{figure*}

  \item The numerical results in Fig.~\ref{fig:eom_phi}(a) show that in the large-radius region, the equation of motion for $\phi$ (\ref{eomP}) is reduced to
  \be A^{-1}e^{\delta}\phi_{,tt}\approx-(A^{-1}e^{\delta})_{,t}\phi_{,t}.\label{phi_tt_1}\ee
  Using Eq.~(\ref{phi_tt_1}) and the numerical results of $|\delta_{,t}|\gg |A_{,t}|$, we have
  \be \phi_{,tt}\approx-\delta_{,t}\phi_{,t}.\label{phi_tt_2}\ee
  Combining Eqs.~(\ref{phi_t}), (\ref{H_t}), and (\ref{phi_tt_2}) and the numerical results of $|\delta_{,t}|\gg H_{,t}$ generates
  \be [H]_{,tt}\approx\frac{\omega\alpha_{,t}}{r}\approx -\delta_{,t}[H]_{,t}.\label{H_tt}\ee
  Namely, we treat $\alpha$ effectively as constant at the first-order accuracy, and the dynamical feature of $\alpha$ begins to take effect since the second-order temporal derivative of $[H(r,t)]$.
  \item As shown in Fig.~\ref{fig:eom_phi}(b), at the late stage of critical collapse, in the large-radius region at which $|t/r|\ll 1$, $|H|\ll 1$, $|H|\ll|\omega\ln r|$, and $|H_{,r}|\ll 1/r$. Therefore, with Eq.~(\ref{phi_sln}), $[H]$ mainly contributes to the temporal derivatives of $\phi$, rather than to the field $\phi$ and its spatial derivatives.
\end{enumerate}

The numerical results show that $C_{1}\approx0.058$, $C_{2}^2+C_{3}^2\approx 1$, and $\varphi_{1}\approx 1.08$. As shown in Figs.~\ref{fig:phi_fit}(a) and \ref{fig:phi_fit}(b), the expressions for $\phi$~(\ref{phi_sln}), $\phi_{,t}$~(\ref{phi_t}), and $\phi_{,tt}$~(\ref{phi_tt_2}) agree well with the numerical results. Note that when we compare the analytic expressions with the numerical results in the large-radius region, the tail part, e.g., the region for $r>1$ in Fig.~\ref{fig:phi_fit}, is excluded.

With Eqs.~(\ref{phi_sln}) and (\ref{phi_t}), one can rewrite Eq.~(\ref{eqAt}) as

\be
\frac{1}{A}\frac{\partial A}{\partial t}=-8\pi r\phi_{,t}\phi_{,r}
\approx C_{4}[H]_{,t}[\sin(2\omega\ln r + 2C_{3}[H] + 2\varphi_{0} + \varphi_{1})-\sin\varphi_{1}],
\ee
where $C_4=4\pi\omega{C_{1}^2}\sqrt{C_{2}^2+C_{3}^2}$. Via integration, we have
\be
\ln A\approx -\frac{C_4}{2C_3}\cos(2\omega\ln r + 2C_{3}[H] + 2\varphi_{0} + \varphi_{1} )- C_{4}\sin\varphi_{1}[H] + C_5.
\ee
Then, using Eq.~(\ref{MS_mass}) and the fact that $|H|\ll 1$, we obtain
\be \frac{m}{r}\approx C_{6}\cos(2\omega\ln r+2C_{3}[H]+2\varphi_{0}+\varphi_{1})+C_7,\label{m_r_sln}\ee
where $C_6\approx e^{C_5}C_{4}/(4C_3)\approx e^{C_5}\pi\omega{C_{1}^2}\sqrt{C_{2}^2+C_{3}^2}/C_{3}$ and $C_7=(1/2)(1-e^{C_{5}})$. As shown in Fig.~\ref{fig:phi_fit}(c), the expression for $m/r$ (\ref{m_r_sln}) is consistent with the numerical results. The fitting results are $C_6=0.013360\pm 0.000009\approx 1/75$, and $C_7\approx 0.065480\pm 0.000007\approx 1/15$.

With Eq.~(\ref{MS_mass}), we can rewrite Eqs.~(\ref{eomA}) and (\ref{eomd}) as
\be m_{,r}=2\pi r^2 A(P^2+Q^2),\label{eq_mr}\ee
\be r\delta_{,r}=\frac{\partial\delta}{\partial\ln r}=-\frac{2}{1-\frac{2m}{r}}m_{,r}.\ee
Then, the solution for $\delta$ can be expressed as
\be
\delta\approx C_8\ln r +\ln\left(1-\frac{2m}{r}\right)+C_9\sin(2\omega\ln r+2C_{3}[H]+2\varphi_{0}+\varphi_{1})+\delta_{0}(t),
\label{delta_sln}
\ee
where $C_8\approx-{2C_7}/(1-2C_7)-2C_{6}^2$ and $C_9\approx-(C_6+8C_{6}C_{7})/\omega$. As shown in Fig.~\ref{fig:phi_fit}(d), the expression for $\delta$ (\ref{delta_sln}) is consistent with the numerical results.

In Ref.~\cite{Price:1996sk}, the quantities $\alpha$ and $\beta$ were treated as constant. The approximate expressions for $X$ and $Y$ obtained in this manner agree well with the numerical results. Then, it was stated that in this circumstance, the spacetime is effectively flat. In fact, $X$ and $Y$ are combinations of the metric functions and derivatives of the scalar field, rather than the scalar field. To check whether the spacetime is effectively flat, it may be more appropriate to directly investigate the behavior of the equation of motion for the scalar field~(\ref{eomP}). As shown in Fig.~\ref{fig:eom_phi}(a), in the large-radius region, Eq.~(\ref{eomP}) is reduced to Eq.~(\ref{phi_tt_1}), which is clearly different from the flat-spacetime form $\phi_{,tt}=r^{-2}(r^{2}\phi_{,r})_{,r}$. Therefore, the spacetime in this region is not effectively flat.

\section{Result II: dynamics in the central region\label{sec:central}}
As shown in Fig.~\ref{fig:eom_phi}(a), in the central region, the absolute values of the terms $(A^{-1}e^{\delta})_{,t}\phi_{,t}$ and $(A^{-1}e^{\delta})_{,r}\phi_{,r}$ in Eq.~(\ref{eomP}) are considerably lower than those of $A^{-1}e^{\delta}\phi_{,tt}$, $Ae^{-\delta}\phi_{,rr}$, and $(2/r)Ae^{-\delta}\phi_{,r}$. Moreover, in this region, $A\approx1$ and $\delta\approx 0$. Consequently, Eq.~(\ref{eomP}) is reduced to the flat-spacetime form
\be \phi_{,tt}\approx \frac{1}{r^2}(r^{2}\phi_{,r})_{,r}.\label{eom_phi}\ee

Regarding Eq.~(\ref{eom_phi}), we discuss the following:
\begin{enumerate}[fullwidth,itemindent=0em,label=(\roman*)]
\item Equation~(\ref{eom_phi}) implies that in the central region, the scalar field $\phi$ evolves almost as in flat spacetime, not experiencing the gravitational effects.

\item Besides critical collapse, we check the evolution of the scalar field in two other types of collapse (dispersion and black hole formation) and obtain similar results as~(\ref{eom_phi}).
\begin{figure}[!htbp]
    \centering
    \includegraphics[width=0.4\textwidth]{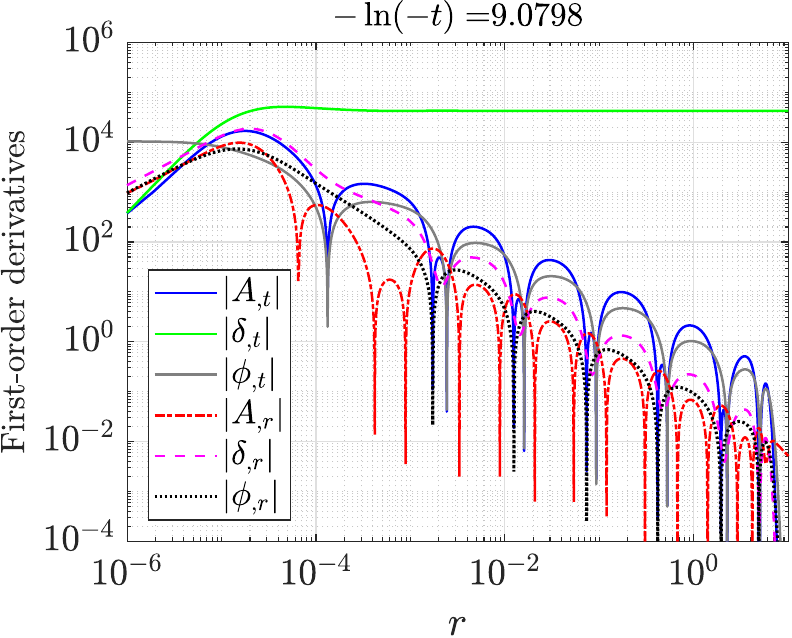}
    \caption{(color online) First-order temporal and spacial derivatives for $A$, $\delta$, and $\phi$. As discussed in Eq.~(\ref{metric_derivatives}), near the center, $A_{,t}\propto r^2$, $\delta_{,t}\propto r^2$, $\phi_{,t}\approx \phi_{0}'(t)+\phi_{2}'(t)r^2$, $A_{,r}\propto r$, $\delta_{,r}\propto r$, and $\phi_{,r}\propto r$.}
    \label{fig:derivatives}
\end{figure}
\item The result~(\ref{eom_phi}) is closely related to the asymptotic behaviors of the metric functions and scalar field near the center. Under the smoothness requirement at the center, the metric functions and scalar field have the following power series expansions near the center~\cite{Zhang:2016kzg}:
\be A\approx 1+A_{2}(t)r^2, \phantom{ddd}\delta\approx\delta_{2}(t)r^2, \phantom{ddd}\phi\approx\phi_0(t) + \phi_{2}(t)r^2.\label{metric_aymptotes}\ee
With Eqs.~(\ref{eomA}), (\ref{eqAt}), (\ref{eomd}), (\ref{MS_mass}), and (\ref{metric_aymptotes}), we obtain the following asymptotic expressions:
\be
\begin{array}{l l}
A_{,t}\approx -16\pi \phi_{,t}\phi_{2}r^2, \hphantom{dddd} \delta_{,t}\approx -4\pi \phi_{,tt}\phi_{,t}r^2, \hphantom{dddd} \phi_{,t}\approx \phi_{0}'(t)+\phi_{2}'(t)r^2,\\
\\
A_{,r}\approx -\frac{8\pi}{3}(\phi_{,t})^2 r, \hspace{1pt}\hphantom{ddddd}\delta_{,r}\approx -4\pi(\phi_{,t})^2 r, \hspace{3pt}\hphantom{ddddd}  \phi_{,r}\approx 2\phi_{2}(t)r,\\
\end{array}
\label{metric_derivatives}
\ee
which are also shown in Fig.~\ref{fig:derivatives}. With Eqs.~(\ref{metric_aymptotes}) and (\ref{metric_derivatives}), we can straightforwardly simplify Eq.~(\ref{eomP}) to (\ref{eom_phi}).

\item It is known that in critical collapse, the Ricci curvature scalar $R$ in the central region is very high and will eventually diverge. This does not contradict the result in Eq.~(\ref{eom_phi}) because Eq.~(\ref{eom_phi}) is caused by the fact that the first-order derivatives of the metric components asymptote to zero. Conversely, as discussed below, the major terms constructing $R$ include the second-order derivatives of the metric components, first-order derivatives of the metric components divided by $r$, and $2(1-A)$ divided by $r^2$, which are very large and will eventually diverge.

    For the metric~(\ref{metric1}), the Ricci curvature scalar can be written as
\be
R=\frac{4A\delta_{,r}}{r} - \frac{4A_{,r}}{r} + 2A\delta_{,rr} + \frac{2(1-A)}{r^2} - A_{,rr} - \frac{A_{,tt}e^{2\delta}}{A^2}
  + 3A_{,r}\delta_{,r} - 2A(\delta_{,r})^2 + \frac{2(A_{,t})^2 e^{2\delta}}{A^3} - \frac{A_{,t}\delta_{,t}e^{2\delta}}{A^2}.
\label{Ricci}
\ee
With Eqs.~(\ref{eomA}), (\ref{eqAt}), (\ref{eomd}), (\ref{MS_mass}), (\ref{metric_aymptotes}), and (\ref{metric_derivatives}), we obtain asymptotic expressions for all the terms on the right-hand side of Eq.~(\ref{Ricci}):
\be
\frac{4A\delta_{,r}}{r}\approx -2D,\hphantom{dd}- \frac{4A_{,r}}{r}\approx\frac{4}{3}D,\hphantom{dd}2A\delta_{,rr}\approx -D,
\hphantom{dd}\frac{2(1-A)}{r^2}\approx- A_{,rr}\approx \frac{D}{3}, \hphantom{dd}\mbox{where}\hphantom{d} D=8\pi(\phi_{,t})^2.
\nonumber
\ee
\be
\begin{split}
-\frac{A_{,tt}e^{2\delta}}{A^2}&\approx 16\pi \phi_{,tt}\phi_{2}r^2,\\[6pt]
3A_{,r}\delta_{,r}&\approx 32\pi^2 (\phi_{,t})^4 r^2,\\[6pt]
-2A(\delta_{,r})^2&\approx -32\pi^2 (\phi_{,t})^4 r^2,\\[6pt]
\frac{2(A_{,t})^2 e^{2\delta}}{A^3}&\approx 512\pi^2 (\phi_{,t})^2 (\phi_{2})^2 r^4,\\[6pt]
- \frac{A_{,t}\delta_{,t}e^{2\delta}}{A^2}&\approx -64\pi^2 \phi_{,tt}(\phi_{,t})^2 \phi_{2} r^4.
\end{split}
\nonumber
\ee
The first five terms are dominant and have the same order of magnitude as $8\pi(\phi_{,t})^2$, and the remaining terms are proportional to $r^2$ or $r^4$ and are negligible.
\end{enumerate}

As shown in Fig.~\ref{fig:eom_phi}(b), the transition region between the central and large-radius regions can be expressed as $r\in [r_1,~r_2]$. At $r=r_{1}$, $|C_{3}H|\sim|\omega\ln r|$; at $r=r_2$, $|C_{3}H_{,r}|\sim \omega/r$.

\section{Summary\label{sec:summary}}
Analytic solutions are important for understanding the dynamics of gravitational collapse. Owing to the complexity of the Einstein equations, seeking the analytic solutions to the equations has been a very challenging issue. In successful circumstances, the equations are usually reduced to ODEs. In critical collapse, the equations remain as PDEs, whereas in the large-radius region and late stage of evolution, the spatial and temporal contributions are separate to some extent. This enables us to obtain approximate analytic expressions for the metric functions and matter field.

The boundary conditions at the center play a key role in the dynamics in the central region. In this region, owing to the boundary conditions, the terms related to the gravitational effects in the equation of motion for the scalar field are negligible, such that the equation is reduced to the flat-spacetime form.

\section*{Acknowledgments}\small
The authors are very thankful to Xiao-Kai He, Junbin Li, and Cheng-Yong Zhang for helpful discussions. JQG is supported by Shandong Province Natural Science Foundation under
grant No.~ZR2019MA068. YH, PPW, and CGS are supported by the National Natural Science Foundation of China (Grant No.~11925503).

%%%%%%%%%%%%%%%%%%%%%%%%%%%%%%%%%%%%%%%%%%%%%%%%%%%%%%%%%%%%%%%%%%%%

\end{document}